\documentclass[fleqn,usenatbib]{mnras}

\usepackage{newtxtext,newtxmath}
 
\usepackage[T1]{fontenc}
\usepackage{ae,aecompl}
\usepackage{dblfloatfix}

\usepackage{graphicx}	
\usepackage{amsmath}	
\usepackage{amssymb}	
\usepackage{academicons}

\usepackage{tikz,xcolor,hyperref}

\definecolor{lime}{HTML}{A6CE39}
\DeclareRobustCommand{\orcidicon}{
	\begin{tikzpicture}
	\draw[lime, fill=lime] (0,0) 
	circle [radius=0.16] 
	node[white] {{\fontfamily{qag}\selectfont \tiny ID}};
	\draw[white, fill=white] (-0.0625,0.095) 
	circle [radius=0.007];
	\end{tikzpicture}
	\hspace{-2mm}
}

\foreach \x in {A, ..., Z}{\expandafter\xdef\csname orcid\x\endcsname{\noexpand\href{https://orcid.org/\csname orcidauthor\x\endcsname}
			{\noexpand\orcidicon}}
}

\newcommand{\ovi}{O\,{\sc vi}}

\newcommand{\ha}{H\,$\alpha$} 
\newcommand{\hb}{H\,$\beta$}
\newcommand{\hc}{H\,$\gamma$}

\newcommand{\neoniv}{[Ne\,{\sc iv}]}

\newcommand{\neoniii}{[Ne\,{\sc iii}]}

\newcommand{\helium}{He\,{\sc i}}
\newcommand{\heliumb}{He\,{\sc ii}}

\newcommand{\oiii}{[O\,{\sc iii}]~5007~\AA}

\newcommand{\oxygeniii}{[O\,{\sc iii}]}

\newcommand{\ironii}{[Fe\,{\sc ii}]}
\newcommand{\ironiii}{[Fe\,{\sc iii}]}

\def\vhel{\ifmmode{V_{{\rm HEL}}}\else{$V_{{\rm HEL}}$}\fi}
\def\vsys{\ifmmode{V_{\rm sys}}\else{$V_{\rm sys}$}\fi}
\def\kms{\ifmmode{~{\rm km\,s}^{-1}}\else{~km~s$^{-1}$}\fi}
\def\vlsr{\ifmmode{v_{\rm lsr}}\else{$v_{\rm lsr}$}\fi}

\title[Five new Galactic symbiotic stars in VPHAS+]{Discovery of five new Galactic symbiotic stars in the VPHAS+ survey}

\author[Stavros Akras et al.]{
Stavros Akras$^{1,2,\orcidA{}}$~\thanks{E-mail: stavrosakras@gmail.com}, Denise R. Gon\c calves$^{3}$, Alvaro Alvarez-Candal$^{4,\orcidA{}}$, Claudio B. Pereira$^{4}$
\\
$^{1}$Instituto de Matem\'{a}tica, Estat\'{i}stica e F\'{i}sica, Universidade Federal do Rio Grande, Rio Grande 96203-900, Brazil\\
$^{2}$Institute for Astronomy, Astrophysics, Space Applications and Remote Sensing, National Observatory of Athens, GR 15236 Penteli, Greece\\
$^{3}$Observat\'{o}rio do Valongo, Universidade Federal do Rio de Janeiro, Ladeira Pedro Antonio 43, 20080-090 Rio de Janeiro, Brazil\\
$^{4}$Observat\'{o}rio Nacional/MCTIC, Rua Gen. Jos\'{e} Cristino, 77, 20921-400 Rio de Janeiro, Brazil}

\date{Accepted XXX. Received YYY; in original form ZZZ}

\pubyear{2020}

\begin{document}
\label{firstpage}
\pagerange{\pageref{firstpage}--\pageref{lastpage}}
\maketitle

\begin{abstract}
We report the validation of a recently proposed infrared selection criterion for symbiotic stars (SySts). Spectroscopic data were obtained for seven candidates,
selected from the SySt candidates of Akras et al. (2019, MNRAS, 483, 5077) by employing the new supplementary infrared selection criterion for SySts in the VST/OmegaCAM Photometric H-Alpha Survey (VPHAS+). Five of them turned out to be genuine SySts after the detection of \ha, \heliumb~and \oxygeniii\ emission lines as well as TiO molecular bands. The characteristic \ovi\ Raman-scattered line is also detected in one of these SySts. According to their infrared colours and optical spectra, all five newly discovered SySts are classified as S-type. The high rate of true SySts detections of this work demonstrates that the combination of the \ha-emission and the new infrared criterion improves the selection of target lists for follow-up observations by minimizing the number of contaminants and optimizing the observing time.
\end{abstract}




\section{Introduction}
Symbiotic stars (SySts) are long-period interacting binary systems composed by a white dwarf and an evolved red giant \citep[see review by][and references therein]{munari2019}. The symbiosis of the two companions lead to a number of physical phenomena which can provide further insights into the interaction and evolution of binary systems, mass transfer process, accretion disk formation, formation of jets and bipolar circumstellar envelopes, nova-like thermonuclear outbursts, flickering, and soft X-ray emission, among others. Besides all these intriguing phenomena, SySts have also been suggested as progenitors of type~Ia supernovae \citep{Munari1992,DiStefano2010,Dilday2012,munari2019}. 

The recently updated catalogues of SySts~\citep{Akras2019c,Merc2019} lists only 257 Galactic and 66 extragalactic members. Since the publication of these two catalogues, three more Galactic SySts were discovered: Hen 3-1678~\citep[][]{Lucy2018}, HBHa 1704-05~\citep[][]{Munari2018,Skopal2019} and Gaia18aen~\citep{Merc2020}. Yet, the  overall population of Galactic SySts is still incompatible with the theoretical predictions \citep[3,000 to 400,000 SySts; ][]{Kenyon1986,Munari1992,Kenyon1993,Magrini2003,Lu2006}. 

A search of SySts in the INT Photometric H-Alpha Survey \citep[IPHAS, ][]{Drew2005} resulted in the discovery of 19 new members \citep[][and references therein]{corradi2008,flores2014}. However, the IPHAS list of SySt candidates turned out to be substantially contaminated with Be stars and young stellar objects (YSOs). \cite{flores2014} refined the IPHAS selection criterion and ended up with a shorter list of only 162 candidates. Despite this refinement, the second follow-up spectroscopic survey of 18 candidates resulted in five new SySts or 27.7 per cent of identification rate, and the authors argued for the need of even more robust selection criteria in order to distinguish SySts from other \ha-emitters in a more efficient way.

\begin{table*}
\begin{center}
\caption{SOAR observing log} 
\begin{tabular}{cccccccccc}
\hline
\hline
\noalign{\smallskip}
VPHAS+ ID &\multicolumn{1}{c}{R.A.} & \multicolumn{1}{c}{Dec.} &
\multicolumn{2}{c}{t$_{\rm exp}$ (sec)} & \multicolumn{1}{c}{Airmass}  & \multicolumn{1}{c}{Other Names} & \multicolumn{1}{c}{Classification} \\
 & \multicolumn{1}{c}{(J2000.0)} & \multicolumn{1}{c}{(J2000.0)} &
\multicolumn{1}{c}{M1} & \multicolumn{1}{c}{M2} & \multicolumn{1}{c}{} & \multicolumn{1}{c}{} & \multicolumn{1}{c}{} \\
\hline
DR2J141301.4-6533201.1  & 14:13:01.4 & -65:33:20 & 540 & 480 & 1.34-1.38 & WRAY 15-1180 & Genuine SySt \\ 
DR2J172830.6-2921241.5  & 17:28:30.6 & -29:21:24 & $-$ & 660 & 1.04 & [KW2003] 21 & emission line star \\ 
DR2J175320.4-2953271.4  & 17:53:20.4 & -29:53:27 & 780 & 540 & 1.06-1.11 & SS 326 & Genuine SySt  \\
DR2J175346.2-2848261.6  & 17:53:46.2 & -28:48:26 & 900 & 780 & 1.16-1.22 & $-$ & Probable SySt\\   
DR2J181123.2-2414301.0  & 18:11:23.2 & -24:14:30 & 1200 & 1200 & 1.47-1.60 & [KW2003] 95 & Genuine SySt \\      
DR2J181154.5-2435361.2  & 18:11:54.5 & -24:35:36 & 960 & 780 & 1.23-1.30 & [KW2003] 97 & Genuine SySt \\ 
DR2J181333.6-2452251.0  & 18:13:33.6 & -24:52:25 & 780 & 540 & 1.11-1.15 & [KW2003] 98 & Genuine SySt \\
\hline
\end{tabular}
\end{center}
\end{table*}

Machine learning and data mining have gained significant popularity in astronomy over the past few years due to the large amount of data provided from several on-going photometric surveys star-galaxy classification  \citep[e.g.][]{Fadely2012,Clarke2019a}; symbiotic stars \citep{Akras2019a}; planetary nebulae \citep{Akras2019b,Iskandar2020}; low-metallicity stars \citep{Whitten2019}; exoplanet transits \citep{Schanche2019}; Herbig Ae/Be and classical Be stars \citep{Vioque2020}; automatic classification of variable stars \citep{Hosenie2020}; automatic determination of stellar temperatures and metallicities \citep{Karnavas2020}, among many others.

\cite{Akras2019a} devised a number of new infrared (IR) selection criterion using a machine learning approach which, in conjunction with the IPHAS optical criterion (\ha-r), can better distinguish SySts from other \ha-sources in the IPHAS. The new IR criterion was also applied to the VST/OmegaCAM Photometric H-Alpha Survey \citep[VPHAS+, ][]{Drew2014}. In total, 72 new SySt candidates were found, whilst up to 90 per cent of the known SySts in both catalogs were recovered.

The spectroscopic follow-up results of the application of the optical+IR selection criteria to identify genuine SySts is the focus of this letter, which demonstrates that the new IR selection criterion is valid and very efficient. This letter is organised as follows: an overview of the observations is described in Section 2. In Section 3, we present the results of this mini-spectroscopic survey and we finish with the conclusions in Section 4.

\section{Observations} 
\label{sec:obs} 
Low-resolution observations were carried out remotely from Rio de Janeiro (Observatório Nacional/MCTIC) in 2019 August 3$^{\rm{rd}}$ and 7$^{\rm{th}}$ using the Goodman High Throughput Spectrograph \citep{Clemens2004} on the 4.1~m telescope at the Southern Astrophysical Research (SOAR) at the Cerro Tololo observatory in Chile. 
The spectrograph was equipped with the red CCD camera with 4096$\times$4112 square pixels of 0.15~arcsec~pixel$^{-1}$. A two-by-two binning was employed in both the spatial and spectral directions to increase the signal-to-noise ratio. The 1.2~arcsec slit width and the 400~lines~mm$^{-1}$ grating were selected, providing a dispersion of 1\AA~pixel$^{-1}$ and a wavelength range from 3000 to 9050~\AA\ split in two parts: (i) the blue part (setup M1) from 3000 to 7050~\AA\ and (ii) the red part (setup M2) from 5000 to 9050~\AA. This configurations allow us to observe the whole optical spectrum and cover all the characteristic optical lines (\ha, \oxygeniii~$\lambda$5007, \heliumb~$\lambda$4686, \ovi~$\lambda$6830, etc.) and the molecular bands (TiO, VO, etc.) crucial for the confirmation of the symbiotic nature of the candidates. The spectra of the seven candidates are displayed in Figs.~\ref{obj1} and \ref{obj2}. 

The exposure times varied from 180 up to 1200~sec depending on the candidates brightness (Table~1). The blocking filter GG-455 in the M2 setup was also used to avoid contamination from the second order. Wavelength calibration was performed using HgAr+Ne calibration lamp, which provides us with a sufficient number of lines. The reduction of the data was carried out using {\sc iraf} standard procedures. Due to the presence of clouds during the observations and poor weather conditions, the spectra are not flux-calibrated. Nevertheless, this does not prevent us from unveiling the nature of the candidates.

\section{Results} 
Seven SySt candidates were selected from \citet{Akras2019a} for this pilot spectroscopic survey. All of them satisfy both the IPHAS criterion \citep[(i),][]{corradi2008} and the new IR criterion \citep[(ii),][]{Akras2019a}: 

\begin{enumerate}
    \setlength\itemsep{0.01em}
    \item \ha\ IPHAS criterion
   \begin{itemize}
    \setlength\itemsep{0.01em}
     \item ({\it r-\ha})$\geq$0.25$\times$({\it r-i} ) + 0.65
   \end{itemize}
   \setlength\itemsep{0.01em}
   \item IR selection criterion for S-type SySts 
   \begin{itemize}
    \setlength\itemsep{0.01em}
    \item J-H$\geq$0.78~\&~0$<$K$_{\rm s}$-W3$<$1.18~\&~W1-W2$<$0.09 or
    \item J-H$\geq$0.78~\&~0$<$K$_{\rm s}$-W3$<$1.18~\&~W1-W2$\geq$0.09~\&~0$<$W1-W4$<$0.92 
   \end{itemize}
   \setlength\itemsep{0.01em}
   \item IR selection criterion among SySts~/~K-giants~/~M-giants
   \begin{itemize}
     \setlength\itemsep{0.01em}
     \item H-W2$\geq$0.206~\&~K$_{\rm s}$-W3$\geq$0.27 
   \end{itemize}
    \setlength\itemsep{0.01em}
   \item IR criterion for S/S+IR/D/D$^{\prime}$ scheme classification
   \begin{enumerate}
    \setlength\itemsep{0.01em}
     \item S-type criterion
      \begin{itemize}
       \setlength\itemsep{0.01em}
        \item K$_{\rm s}$-W3$<$1.93~\&~W3-W4$<$1.46
      \end{itemize}
      \setlength\itemsep{0.01em}
      \item S+IR-type criterion 
      \setlength\itemsep{0.01em}
      \begin{itemize}
      \setlength\itemsep{0.01em}
       \item K$_{\rm s}$-W3$<$1.93~\&~W3-W4$\geq$1.46 or
       \item K$_{\rm s}$-W3$\geq$1.93~\&~H-W2$<$2.72 
      \end{itemize}
      \setlength\itemsep{0.01em}
      \item D-type criterion
      \setlength\itemsep{0.01em}
      \begin{itemize}
        \setlength\itemsep{0.01em}
        \item K$_{\rm s}$-W3$\geq$1.93~\&~H-W2$\geq$2.72~\&~W3-W4$<$1.52
      \end{itemize}
      \setlength\itemsep{0.01em}
      \item D$^{\prime}$-type criterion
      \setlength\itemsep{0.01em}
      \begin{itemize}
        \setlength\itemsep{0.01em}
        \item K$_{\rm s}$-W3$\geq$1.93 and H-W2$\geq$2.72 and W3-W4$\geq$1.52
      \end{itemize}
   \end{enumerate}
\end{enumerate}

For the confirmation of a candidate as a genuine SySt, specific optical lines must be detected such as \ha, \oxygeniii~$\lambda$5007, \heliumb~$\lambda$4686, and \ovi~6830, commonly found in the spectra of SySts. These lines originate from the highly ionized circumstellar envelope illuminated by the strong UV field of the white dwarf. Moreover, a number of molecular bands like TiO and VO associated with the cold red giant, are also present in the spectra. The latter features are easily perceptible in the S-type SySts but not in the heavily dusty D-types \citep[see][]{munari2002}.

\setcounter{figure}{0}
\begin{figure*}
\includegraphics[scale=0.3]{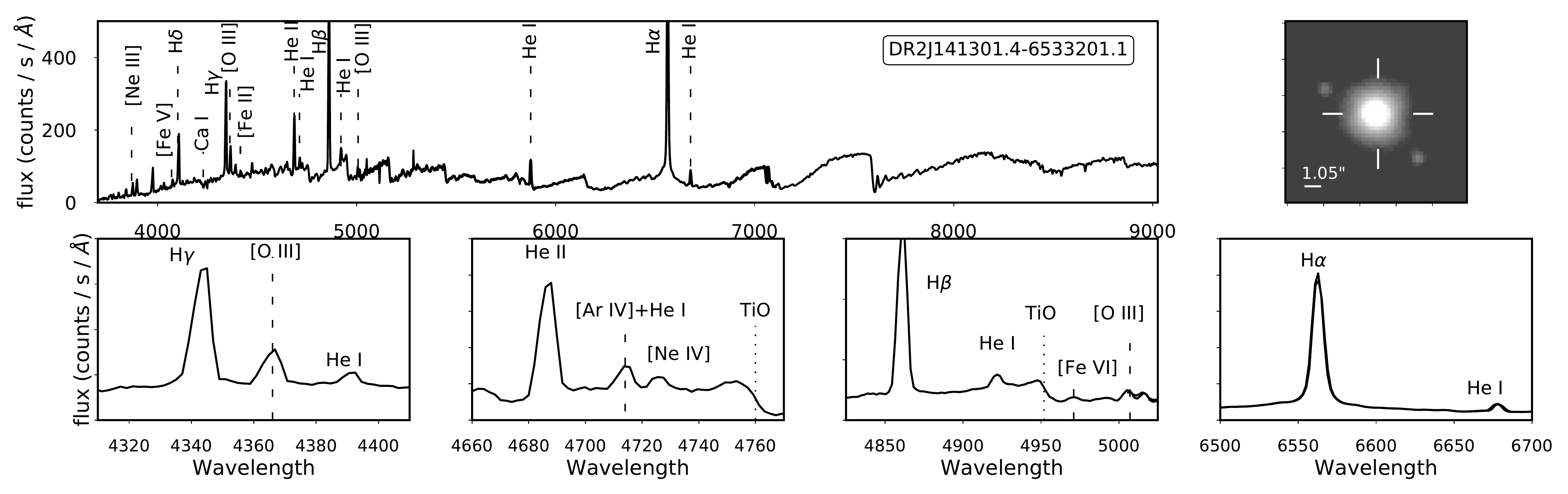}
\includegraphics[scale=0.3]{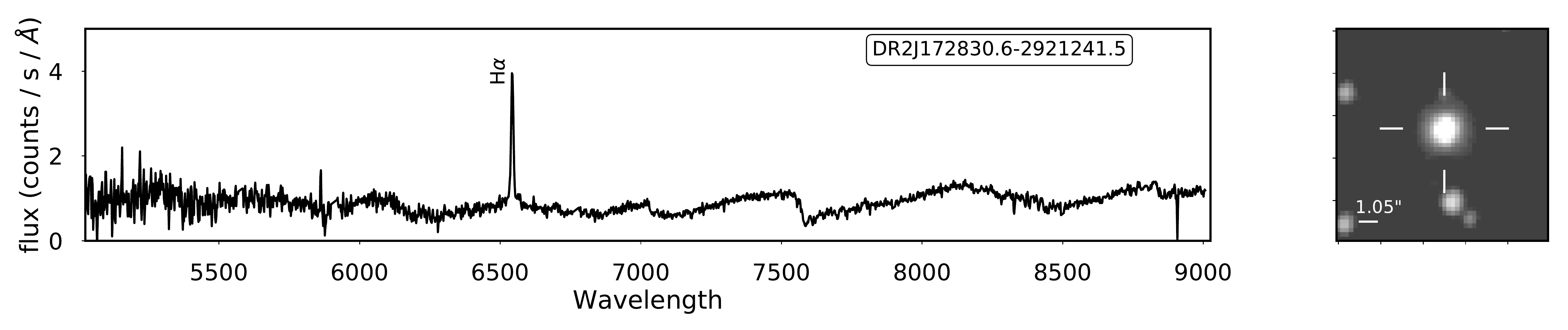}
\includegraphics[scale=0.3]{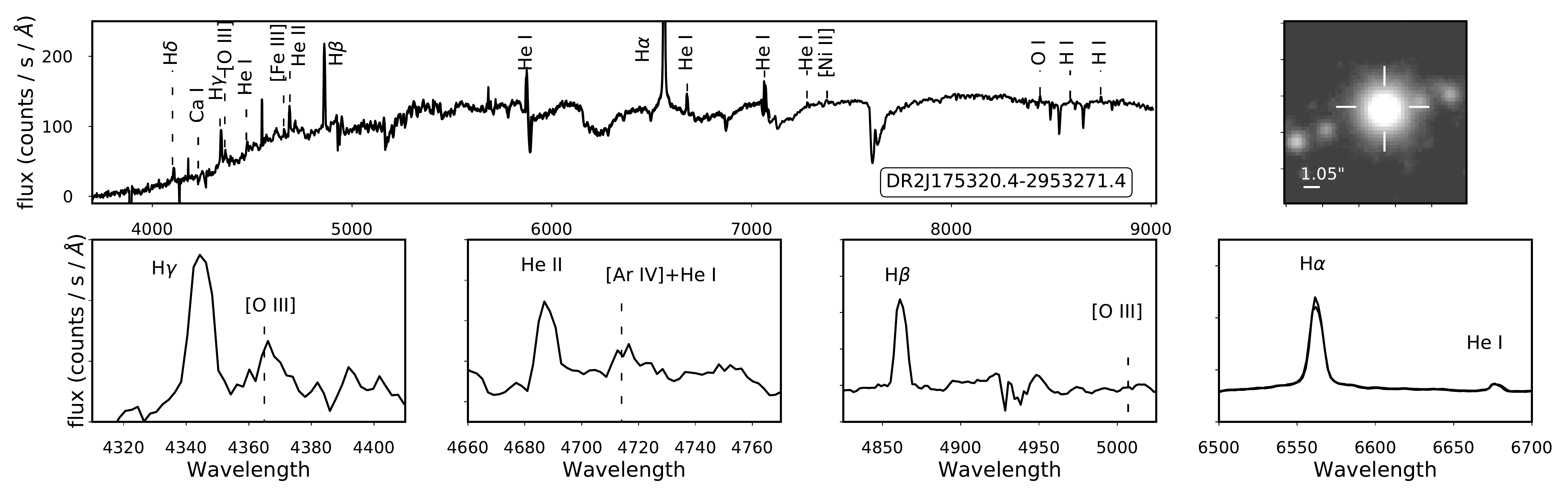}
\includegraphics[scale=0.3]{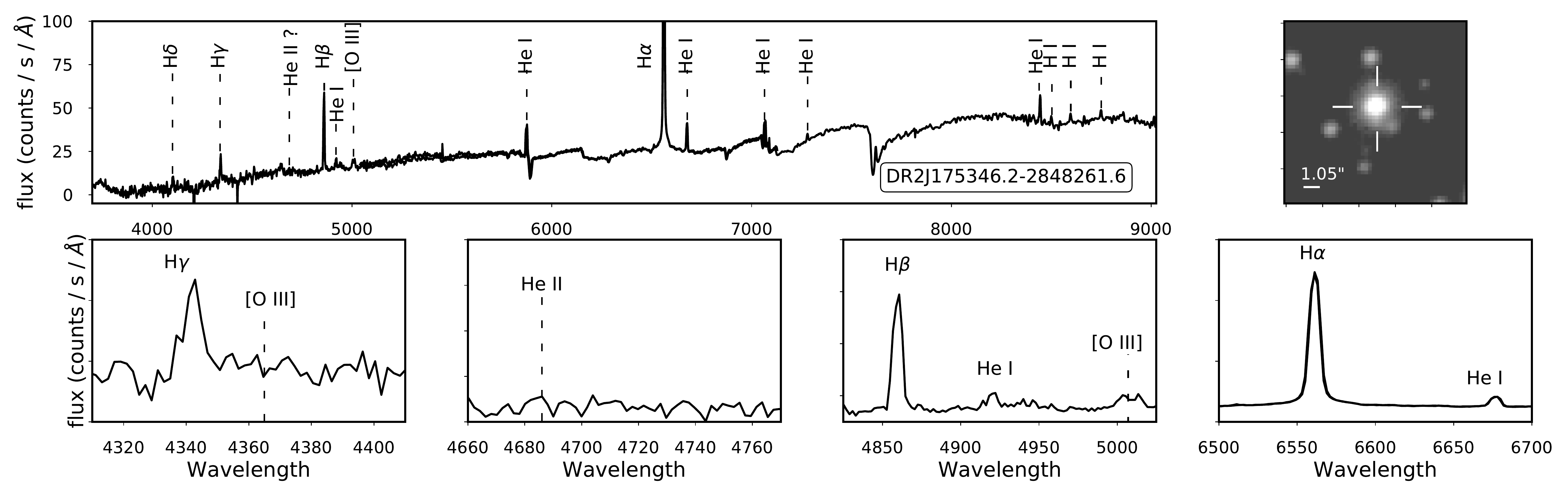}
\caption[]{Low resolution spectrum of the VPHAS+ SySts selected from \citep{Akras2019a}. Top left panels show the observed spectra. The top right panels display the VPHAS+ \ha\ images of the sources in a logarithmic scale. North is up, east to the left. Bottom panels zoom in the \hc~and \oxygeniii~4363\AA\ lines, \heliumb~4686\AA~line, \hb~and\oiii\ lines, and \ha\ emission lines. The order of the sources is the same as in Table~1.}
\label{obj1}
\end{figure*}

\begin{figure*}
\includegraphics[scale=0.3]{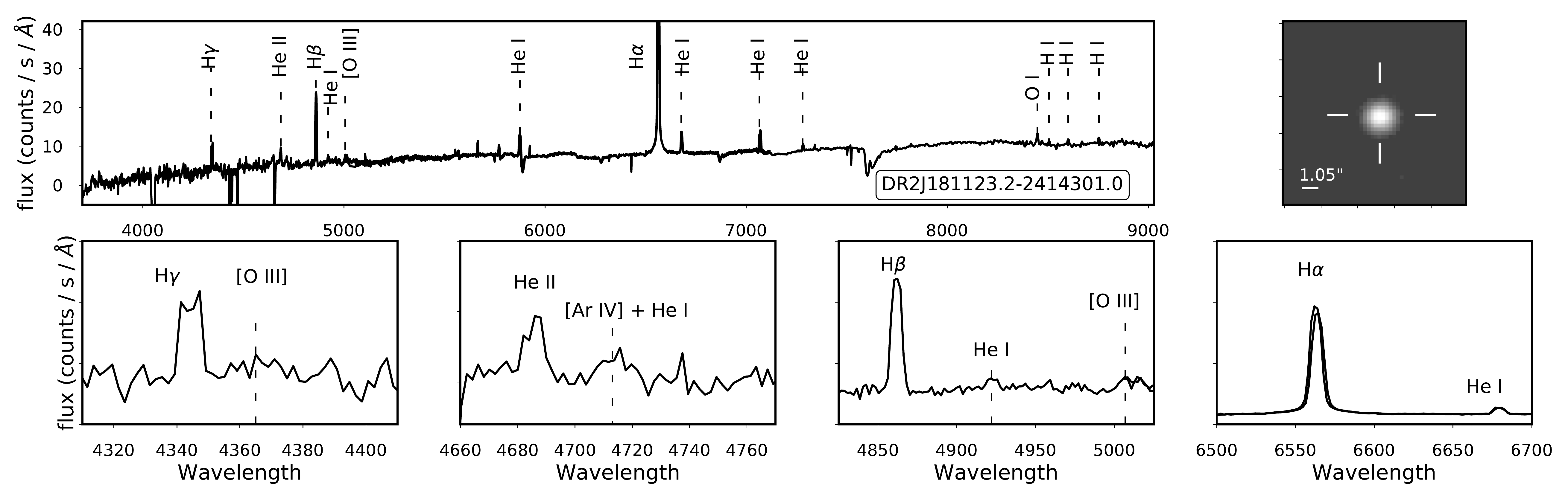}
\includegraphics[scale=0.3]{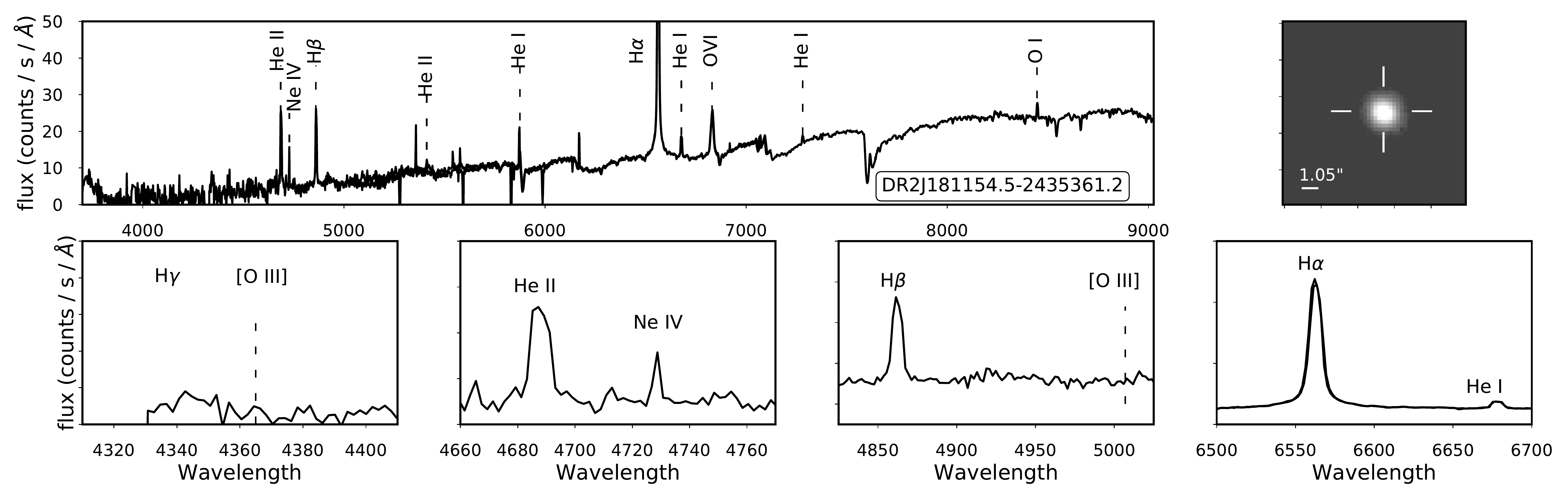}
\includegraphics[scale=0.3]{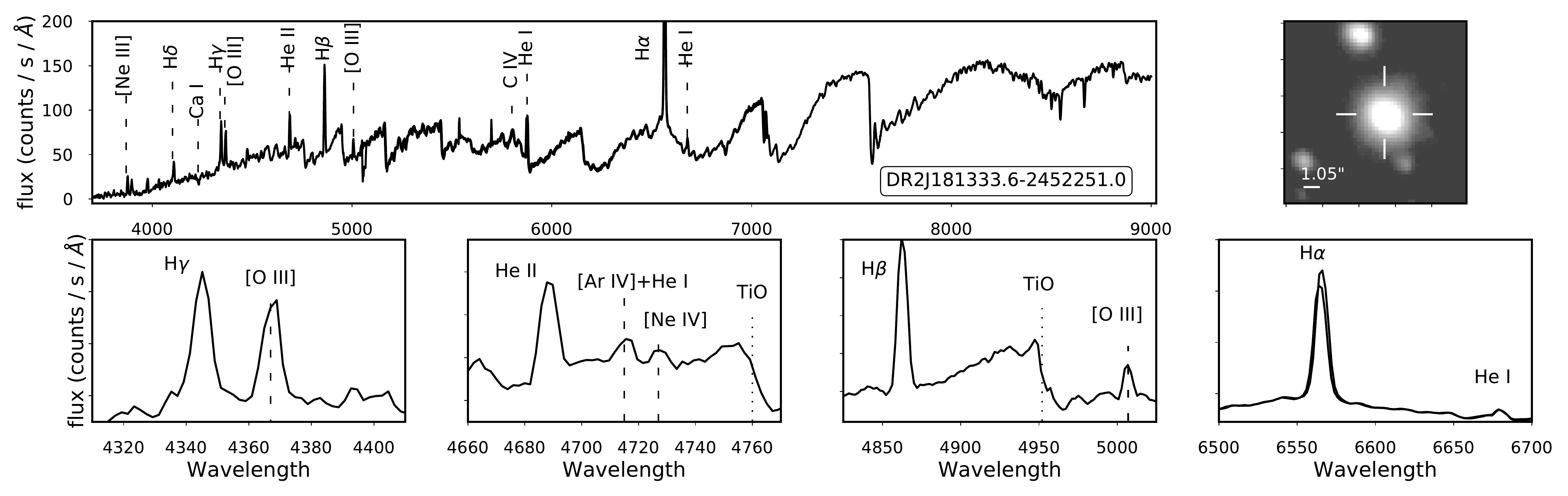}
\caption[]{The same as Figure~1.}
\label{obj2}
\end{figure*}

Most of these features are detected in our candidates, thus confirming their symbiotic nature (Figs.~\ref{obj1} and \ref{obj2}). In particular, strong \ha\ emission line is found in all the candidates, as it was expected based on their positions in the IPHAS {\it r-\ha} versus {\it r-i} diagnostic diagram. The recombination \helium\ and \heliumb\ lines are also detected. Five of the candidates clearly exhibit the \heliumb~4686\AA\ line. \oxygeniii~5007\AA\ and 4363\AA\ emission lines are detected in four and three candidates, respectively. Of particular interest is the candidates DR2J175320.4-2953271.4 in which the \oxygeniii~5007\AA\ line is not detected, while we detected the \oxygeniii~4363\AA~line. The same has been found in other known SySts \citep[e.g. Hen 3-863, AG Peg, AS 327 , see ][]{Luna2005} and all of them have extreme high electron densities (10$^8$ cm$^{-3}$). Such high densities lead to the attenuation of the \oxygeniii~5007\AA\ emission relative to the 4363\AA\ one and to the different loci of SySts and planetary nebulae in the \oxygeniii~5007/\hb~ versus \oxygeniii~4364/\hc~ diagnostic diagram \citep[][]{GutierrezMoreno1995,Clyne2015}.
Emission lines such as \neoniii, \neoniv, \ironii, and \ironiii\ are also detected implying a high excitation circumstellar envelope. The \ovi\ Raman-scattered line at 6830\AA, present in principle only in SySts \citep[][]{Allen1980,Akras2019c,Angeloni2019} is also detected in one of our candidates (DR2J181154.5-2435361.2).

Four of the candidates clearly show TiO molecular bands associated with the presence of a red giant with an M spectral type assigned. The Ca~I~4227\AA~absorption line is also detected in three of them with TiO bands in 4760 and 4955~\AA. The absence or very weak TiO bands of DR2J181123.2-2414301.0 implies a late K or early M type red giant.

Two of the newly classified SySts (DR2J141301.4-6533201.1, DR2J181123.2-2414301.0) and the probable one (DR2J175346.2-2848261.6) display a seemingly double-peak \oxygeniii~$\lambda$5007 line. Notice that the second peak is found at 5015\AA\ and it corresponds to the \helium~ recombination line.

Overall, five out the seven candidates are classified as genuine SySts and one as probable due to the absence of the \heliumb~4686\AA\ and weak \oxygeniii~5007\AA\ emission lines (Table~1). It should be noted that the candidate DR2J175346.2-2848261.6 is also included in the recent work by \cite{Vioque2020}, who employed a machine learning approach to identify new YSOs (Herbig Ae/Be or T Tauri types) and Be stars in GAIA DR2. The probability of DR2J175346.2-2848261.6 to be a YSO or a Be star is as low as 0.04 and 0.003~percent, respectively. Thus, we decided to classify this source as a probable SySt. The remaining candidate (DR2J172830.6-2921241.5) exhibits a spectrum of an early M-type star with clear TiO bands and strong \ha\ emission typically of young stellar objects, but its probability to be a YSO (or Be star) is very low \citep[0.013 (0.003) percent, ][]{Vioque2020}.

SySts are classified either as stellar (S- and S+IR-type) or dusty (D- and D$^{\prime}$-type) based on their spectral energy distribution (SED) \citep[e.g., ][]{flores2014,Akras2019c}. Instead of their SEDs, we make use of the criterion (iv) devised by \cite{Akras2019a}. The K$_{\rm s}$-W3 colour is the main criterion that separate the S-type from the rest of them. All the newly discovered SySts have K$_{\rm s}$-W3$<$1.93 and they are classified as S-type (iv-a). The candidate SySt (DR2J175346.2-2848261.6) exhibits W3-W4$>$1.46 and satisfies the criterion for S+IR-type (iv-b) -- SySts with S-type SED and a potential excess in W3 and/or W4 bands \citep[see ][]{Akras2019c}.

This spectroscopic study yields to an identification rate of 71 percent which is very close to the expected rate of the IR selection criterion \citep{Akras2019a}. Our results also validate the use of the new IR criterion for S-type SySts, and they should be used by the community in order to get higher confidence regarding the symbiotic nature of potential candidates, as well as to select better targets for follow-up spectroscopic surveys and thus optimise the use of observing facilities.

Besides the new discoveries presented here, it is worth mentioning that \cite{Lucy2018} reported the discovery of 10 candidate SySts using the SkyMapper \citep{Keller2007,wolf2018}. Seven of them satisfy the new IR selection criterion of SySts (ii). Two of these seven candidates, Hen~3-1768 and GSC 09276-00130, were observed during the verification tests of the RAMSES-II project \citep{Angeloni2019} and a Raman-scattered \ovi~$\lambda$6830 line excess was reported for both objects. Follow-up spectroscopic observation of Hen~3-1768 shows clearly the presence of the \ovi~, \heliumb~and \ha\ emission lines \citep{Lucy2018}. GSC~09276-00130, on the other hand, was found to be an M-type giant \citep[][]{Angeloni2019}, without satisfy the IR criterion from the SySts/K-giant/M-giant model (iii). This result demonstrates that the additional criteriin from the SySts/K-giant/M-giant model can further distinguish single K- and M-type giants from SySts. 

The second newly discovered Galactic SySt (HBHA~1704-05) reported by \cite{Munari2018}, satisfies the IR criterion of SySt (ii) as well as those criterion from the SySts/K-giant/M-giant model (iii). The spectroscopic detection of the \heliumb\ and \ovi\ lines have confirmed its symbiotic nature \citep{Skopal2019}.

Finally, \cite{Merc2020} have recently presented the discovery of a new SySt (Gaia18aen) in the GAIA catalog. This third Galactic SySt also satisfies both selection criteria (ii \& iii) providing an additional validation.

\section{Discussion and Conclusions}
The results from the spectroscopic pilot survey of seven candidate SySts in the VPHAS+ survey selected from \cite{Akras2019a} were presented. All the candidates satisfy both the optical IPHAS criterion and the new IR selection criterion for SySts. Five of them are confirmed genuine SySts based on the detection of various characteristic optical lines and molecular bands and one is actually a probable SySt. Moreover, based on the IR criterion all newly discovered SySts were classified as S-type and the probable as an S+IR-type. 

The discovery of these five new genuine SySts provided the necessary validation of the IR selection criterion, with a detection rate of 71 percent. Hence, the optical IPHAS criterion of \ha-emission in conjunction with the new IR selection criterion for SySts are proven to be highly efficient in the discrimination of SySts over their \ha-mimics and increase the detection rates of new discoveries. We thus recommend the community to use these new supplementary IR criterion to further support any new discovery of SySts as well as to select targets for follow-up spectroscopic surveys significantly less contaminated by Be stars or YSOs.

Based on the number of SySt candidates in the VPHAS+~DR2, which covers only 24 per cent of the entire footprint of the survey, and the efficiency of the new IR criterion, many new genuine SySts are expected to be discovered in the next VPHAS+ data releases. We plan to continue the spectroscopic follow-up of IPHAS and VPHAS+ SySt candidates or any candidates obtained from other photometric surveys which provide the necessary information about \ha\ emission, like the Javalambre Physics of the Accelerating Universe Astrophysical Survey \citep[J-PAS, ][]{Benitez2014}, Javalambre Photometric Local Universe Survey \citep[J-PLUS, ][]{Cenarro2019}, and the Southern Photometric Local Universe Survey \citep[S-PLUS, ][]{MendesdeOliveira2019}.

\section*{Acknowledgements}
We thank the anonymous referee for his/her comments which help us to improve the manuscript. AAC acknowledges support from CNPq (grants  304971/2016-2  and 401669/2016-5) and the Universidad de Alicante (contract UATALENTO18-02). Based on observations obtained at the Southern Astrophysical Research (SOAR) telescope, which is a joint project of the Minist\'erio da Ciencia, Tecnologia, Inova\c{c}\~oes e Comunica\c{c}\~oes do Brasil (MCTIC/LNA), the U.S. National Science Foundation's National Optical Astronomy Observatory (NOAO), the University of North Carolina at Chapel Hill (UNC), and Michigan State University (MSU).). 

\section*{DATA AVAILABILITY}
The spectroscopic data underlying this article are available at the SOAR Archive (http://ast.noao.edu/data/archives). The photometric data were obtained from publicly available catalogues.






\bibliographystyle{mnras}  
\bibliography{references}   


\bsp	
\label{lastpage}
\end{document}